\documentclass[
	aps,
    pra,
    superscriptaddress,
    groupedaddress,
    twocolumn,
    a4paper,
    floatfix,
    raggedbottom,
    longbibliography
]{revtex4-1}

\usepackage[american]{babel}
\usepackage[utf8x]{inputenc}

\usepackage{grffile} 

\usepackage{newtxtext,newtxmath}
\usepackage{microtype}

\usepackage{amsmath}
\usepackage{amssymb}
\usepackage{bm}
\usepackage{mathtools}
\usepackage{braket}

\usepackage{graphicx}
\usepackage[svgnames]{xcolor}

\usepackage{siunitx} 

\usepackage{ragged2e}
\usepackage{array}
\usepackage{tabularx}
\usepackage{booktabs}

\definecolor{cset-aps-blueberry}{RGB}{28,128,158}
\definecolor{cset-aps-blue}{RGB}{46,44,184}
\definecolor{cset-aps-turquoise}{RGB}{0,67,88}
\definecolor{cset-aps-limegreen}{RGB}{190,219,67}
\definecolor{cset-aps-green}{RGB}{31,138,112}
\definecolor{cset-aps-yellow}{RGB}{255,225,25}
\definecolor{cset-aps-orange}{RGB}{253,116,0}
\definecolor{cset-aps-red}{RGB}{219,0,43}

\usepackage{tikz}

\usepackage{pgfplots}
\pgfplotsset{%
    every axis legend/.append style={%
        cells={anchor=west},
        at={(0.96,0.04)},
        anchor=south east,
        font=\scriptsize,
        },
    every axis/.append style={%
        yticklabel style={%
            /pgf/number format/fixed zerofill,
            /pgf/number format/precision=2},
        },
    width= \textwidth,
    height=8cm,
    xmajorgrids=true,
    xminorgrids=false,
    minor x tick num=1,
}
\mathtoolsset{showonlyrefs=true}

\newcommand{\Secref}[1]{Sec.~\ref{#1}}
\newcommand{\Eq}[1]{Eq.~\eqref{#1}}
\newcommand{\Eqs}[1]{Eqs.~\eqref{#1}} 
\newcommand{\Fig}[1]{Fig.~\ref{#1}} 
\newcommand{\App}[1]{Appendix~\ref{#1}}

\usepackage{hyperref}
\hypersetup{%
    colorlinks=true,
    linkcolor={cset-aps-red},
    linkbordercolor={cset-aps-red},
    filecolor={cset-aps-orange},
    filebordercolor={cset-aps-orange},
    citecolor={cset-aps-blue},
    citebordercolor={cset-aps-blue},
    urlcolor={cset-aps-green},
    urlbordercolor={cset-aps-green},
    menucolor={cset-aps-limegreen},
    menubordercolor={cset-aps-limegreen},
    breaklinks=true,
    pdfborderstyle={/S/U/W 2},
    pdfpagemode=UseOutlines,
    pdfstartpage={1},
}

\begin{document}

\title[Title]{Perturbative operator approach to high-precision light-pulse atom interferometry}
\collaboration{Published as
    \href{https://doi.org/10.1103/PhysRevA.101.053615}
    {Physical Review A \textbf{101}, 053615 [2020]}}


\author{Christian Ufrecht}
\email{christian.ufrecht@gmx.de}
\author{Enno Giese}

\address{\vspace{0.2cm}Institut f{\"u}r Quantenphysik and Center for Integrated Quantum Science and Technology (IQ\textsuperscript{ST}), Universit{\"a}t Ulm, Albert-Einstein-Allee 11, D-89069 Ulm, Germany}

\begin{abstract}
 Light-pulse atom interferometers are powerful quantum sensors, however, their accuracy for example in tests of the weak equivalence principle  is limited by various spurious influences like stray magnetic  fields or blackbody radiation. Improving the accuracy  therefore requires a  detailed  assessment of the size of such deleterious effects. Here, we present a systematic operator expansion to obtain phase shift and contrast  analytically in powers of a perturbation potential. The result can
either be employed for  robust straightforward order-of-magnitude  estimates or for rigorous calculations. Together with general conditions for the validity of the approach, we provide a particularly useful formula for the phase including wave-packet effects.
\end{abstract}

\maketitle

\section{Introduction}
Since their first implementation \cite{FirstLPI} in 1991  the accuracy of light-pulse atom interferometers has been improved considerably, which led to high-precision  applications in gravimetry \cite{DroppingAtoms,Syrte}, gradiometry \cite{Gradiometrie1,Gradiometrie3,Gradiometrie4}, tests of fundamental physics such as the weak equivalence principle \cite{WEP1,WEP3,WEP4,WEP5}, measurements of the fine-structure constant \cite{FineStructure1,FineStructure4} and proposals for gravitational-wave detection \cite{GravitationalWaves}. However, as the accuracy is pushed further, an increasing number of formerly negligible influences such as magnetic field gradients, blackbody radiation inducing a spatially-dependent a.c.~Stark shift \cite{BBR2, BBR1}, or gravitational fields of the laboratory environment have to be included into the error budget.
In this article we present a systematic approach to account for such spurious effects.

Although phase shifts caused by the corresponding in generalanharmonic potential shifts can be small, they might nevertheless be non-negligible which calls for a systematic perturbative approach.
Over the years, several powerful analytic methods for the calculation of phase and contrast of light-pulse atom interferometers have been developed based on the Feynman path integral \cite{Tannoudji,Borde2}, descriptions in phase space \cite{Dubetsky,InterfaceGravity} as well as in the form of representation-free descriptions on the operator level for linear gravity \cite{RedshiftSchleich}, path-independent quadratic Hamiltonians \cite{Marzlin,Kleinert} or within a local-harmonic \cite{Hogan, OvercommingLoss} approximation. Within the latter approach, it was also possible to obtain wave-packet effects to lowest order \cite{ZellerPhd}.
However, these methods are either applicable to at most quadratic potentials or lack a comprehensive discussion of consistency in the case of more general applications.
 
In this article we derive a systematic perturbative description for phase and contrast including effects due to wave-packet dynamics based on two formal series, the Magnus \cite{ Magnus2, Magnus1,Ufrecht, QuantumInformation} and the cumulant \cite{ Cumulants2} expansion. They have already been applied in the context of light-pulse atom interferometry \cite{Ufrecht} to determine the quality of magnetic shielding \cite{Magneticshielding} and to calculate relativistic effects for interferometric redshift tests~\cite{Redshift}. Recently, the Magnus expansion was furthermore used to take into account the effect of finite pulse duration~\cite{ Bertoldi}. In this article we extend the work of Ref.~\cite{Ufrecht} with particular emphasis on general conditions for the validity of the approach, characterizing the magnitude of the perturbations. 

In \Secref{Branch-dependent description} we outline our main results and put them into context. Subsequently, in \Secref{Perturbative treatment} we introduce our  path-dependent model and derive the perturbative expansion. The conditions under which our method is valid will be discussed in \Secref{Validity conditions}. Finally in  \Secref{Gravitationalpotential}, as a simple example, we apply the formalism to the cubic potential appearing in the Taylor expansion of the gravitational potential of Earth.

\section{Branch-dependent description}
\label{Branch-dependent description}
\begin{figure}
	\begin{center}
		\includegraphics[width=\linewidth]{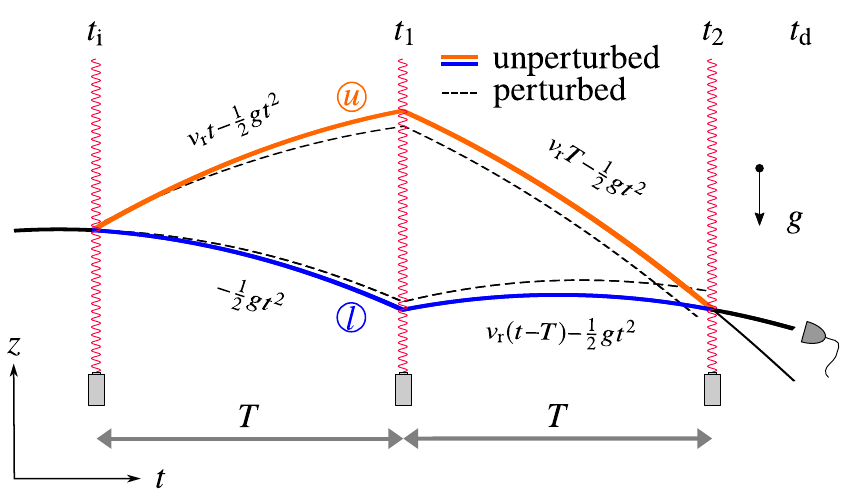}
		\caption{\textit{Mach-Zehnder gravimeter}. Initially released from an atomic trap, the free fall of the atoms  is interrupted by a $\pi/2$ pulse at $t=t_\mathrm{i}=0$ to split the wave packet into two components transferring a momentum  of $\hbar k$ to one of them, which gains  the recoil velocity $v_\mathrm{r}=\hbar k/m$. After a time $T$ both components are redirected by a $\pi$ pulse at $t_1=T$ and finally recombined by a second $\pi/2$ pulse at $t_2=2T$. At the detection time $t_\mathrm{d}$ the number of particles is measured at one of the exit ports, which forms an interference pattern dependent on the relative phase accumulated between the two branches.
		The unperturbed branches (thick solid lines) are determined by the analytic expressions shown next to them, where $g$ is the local gravitation acceleration. The perturbation $V^{(\alpha)}$ slightly disturbs the atoms, leading to the deviating branches of the actual interferometer (thin dashed lines). 
		}
		\label{fig:MZ}
	\end{center}
\end{figure}

Light-pulse atom interferometers consist of a sequence of light pulses which coherently split the initial  wave packet and subsequently direct the  atoms along the two branches of the interferometer.
After recombination by a  final laser pulse, the number of atoms at each exit port displays an interference pattern from which the relative phase accumulated between the  branches of the interferometer can be inferred. In this work we assume that the Hamiltonian describing the motion through the interferometer can be decomposed into a dominant part (linear gravity, laser pulses) and a weak perturbation (e.g.~gravity gradients, blackbody radiation, etc.).
As illustrated by  the Mach-Zehnder (MZ) gravimeter shown in \Fig{fig:MZ}, 
the two branches of the interferometer are mainly caused by the dominant part of the Hamiltonian (thick solid lines in the figure). These trajectories are only slightly disturbed  by the perturbation potentials  (leading to the thin dashed lines in the figure),  which can in general be different for each branch and time dependent. The hypothetical interferometer sequence with vanishing perturbation will be referred to in the following as \textit{unperturbed interferometer}. 

Neglecting wave-packet effects, it has been shown \cite{Tannoudji, Hogan,  ZellerPhd, Ufrecht} that the phase $\phi$ of such an
 interferometer can be obtained from the classical trajectories through
\begin{equation}
\label{Pathintegral}
\phi=\Delta S/\hbar+\phi_\mathrm{s}\,,
\end{equation}
where  $\Delta S=S^{(u)}-S^{(l)}$ is the classical action difference
between the upper and lower branch (superscripts $u$ and $l$)  in combination with a separation phase $\phi_\mathrm{s}$  in case the classical trajectories do not coincide upon detection. In this case we refer to the interferometer  as \textit{open}. The calculation of the phase via \Eq{Pathintegral} therefore consists of (i) solving the differential equation for the classical trajectories including all perturbations and (ii) evaluating the action difference.
In the case of weak perturbing potentials, however,  this approach is unfavorable for the following reasons:
In general, analytic expressions for the trajectories including the perturbation do not exist. Therefore, the trajectories can be solved iteratively to desired order in the perturbation \cite{Hogan} and are then  substituted into \Eq{Pathintegral}, possibly resulting in cumbersome expressions, whereas  a direct perturbative expansion in powers of the weak perturbing potential would be much more  convenient. 
Furthermore, if the perturbation  is only available numerically, in an integration  of \Eq{Pathintegral}  one has to account for both the dominant contribution   as well as the perturbation, which can be difficult numerically since they likely  differ in size by multiple orders of magnitude.
Finally, the validity of \Eq{Pathintegral} is premised on negligible wave-packet effects. However,
in the presence of anharmonic perturbation  potentials the two components of the wave packet will experience different local expansion dynamics along the branches, leading to a slight mismatch and therefore   resulting in  additional phase contributions upon detection. Consequently, it is \textit{a priori} not obvious if these phases are negligible compared to those of \Eq{Pathintegral} originating  from the perturbation.
The approach presented in this article is based on a systematic operator expansion derived from a full quantum-mechanical description of the interferometer to overcome these problems. It allows formulating conditions for its validity determining  exactly when wave-packet effects are negligible or, in turn, calculating their value to desired accuracy.

Denoting the perturbation potential on the upper and lower branch by  ${V}^{(u)}$ and ${V}^{(l)}$, respectively, we will find for a closed unperturbed interferometer sequence
\begin{align}
 \phi=\phi_0&-\frac{1}{\hbar}\oint \mathrm{d} t\, V(t)\\
 \label{simplifiedFormula}
 &-\frac{1}{2\hbar}\oint\! \mathrm{d} t\, V_{ij}(t)\,\langle\hat{\overline{{r}}}_i(t)\hat{\overline{{r}}}_j(t)\rangle
\end{align}
for the phase including the leading-order phase shifts from the perturbation, where repeated indices are summed over.
 In \Eq{simplifiedFormula} we defined 
 $\phi_0$ as the phase of the unperturbed interferometer which can be calculated for example with the general formula provided in Ref.~\cite{TwinPaper}.
 The perturbation potential ${V}(t)=V^{(\alpha)}(\bm{r}^{(\alpha)}_0(t))$ might be different on the upper ($\alpha=u$) and lower branch ($\alpha=l$) and is evaluated at the \textit{unperturbed trajectories} $\bm{r}^{(\alpha)}_0(t)$.
The integrals run from the initial time $t_\mathrm{i}$, where the atoms are released from the trap, up to the detection time $t_d$ on the upper branch and return along the lower branch back to $t_\mathrm{i}$.
We stress that the perturbation potential and its second derivative   $V_{ij}=\partial_i\partial_j V$ are evaluated at the two unperturbed trajectories, which obviates the solution of a possibly involved differential equation.
The initial conditions for the trajectories are $\bm{r}_0(t_\mathrm{i})=\langle \hat{\bm{r}}\rangle$ and $\dot{\bm{r}}_0(t_\mathrm{i})=\langle \hat{\bm{p}}\rangle/m$, where the expectation value is taken with respect to the initial wave packet.
In \Eq{simplifiedFormula} we furthermore defined the operator describing the free evolution of the wave packet
\begin{equation}
\label{WavePacket}
    \hat{\overline{\bm{r}}}(t)=\hat{\bm{r}}-\langle\hat{\bm{r}}\rangle+\frac{\hat{\bm{p}}-\langle\hat{\bm{p}}\rangle}{m}t\,.
\end{equation}
Consequently, $\langle\hat{\overline{{r}}}_i(t)\hat{\overline{{r}}}_j(t)\rangle$, where the indices label the component of the vectors, provides a measure for the width of the wave packet for vanishing perturbation as detailed further below. 
Solving the classical trajectories and the integral in  \Eq{Pathintegral} to first order in the perturbation, the first line in \Eq{simplifiedFormula} can also be derived \cite{TheoremMatterWave} directly from   \Eq{Pathintegral}. The second line, however, describes  wave-packet effects which are not taken into account by \Eq{Pathintegral}. 
In neutron interferometry small perturbations can also be taken into account to first order in a WKB-like treatment \cite{Greenberger}.

\section{Perturbative treatment}
\label{Perturbative treatment}
We now derive \Eq{simplifiedFormula} from a full quantum-mechanical description.
In order to calculate the phase measured by a light-pulse atom interferometer, one has to start from the multi-level  Hamiltonian including the virtual states necessary for the diffraction  process. However, after adiabatic elimination of the auxiliary states \cite{AdiabaticElimi, DoubleBragg}, neglecting atom-atom interactions, and assuming infinitely short laser pulses, the evolution can be reduced to a branch-dependent description \cite{RedshiftSchleich, Ufrecht} in which the phase $\phi$ and contrast $C$ of the interferometer after projection on one exit port is defined by the expectation value   of the \textit{overlap operator}
\begin{equation}
\label{OverlapOperator}
\langle\hat{U}^{(l)\dagger}\hat{U}^{(u)}\rangle=C\mathrm{e}^{\mathrm{i}\phi}
\end{equation}
with respect to the initial wave function, where $\hat{U}^{(\alpha)}$ generates the time evolution along branch $\alpha$. In this article we assume that the Hamiltonian for each branch allows the decomposition
\begin{equation}
\label{Hamiltonian}
\hat{H}^{(\alpha)}=\hat{H}_0^{(\alpha)}+V^{(\alpha)}(\hat{\bm{r}},t)\,.
\end{equation}
The perturbation potential $V^{(\alpha)}(\hat{\bm{r}},t)$ only slightly disturbs the dominant part of the evolution which is caused by the unperturbed Hamiltonian 
\begin{equation}
\label{UnperturbedHamiltonian}
\hat{H}_0^{(\alpha)}=\frac{\hat{\bm{p}}^2}{2m}+mg\hat{z}+ V_\mathrm{em}^{(\alpha)}( \hat{\bm{r}},t)
\end{equation}
 describing the motion of an atom with mass $m$ in the linear gravitational field, where $g$ is the local gravitational acceleration.  The interaction with the laser pulses is modeled by the potentials
 \begin{equation}
 V_\mathrm{em}^{(\alpha)}( \hat{\bm{r}},t)=-\hbar \sum_\ell[\bm{k}^{(\alpha)}_\ell \hat{\bm{r}}+\varphi_\ell^{(\alpha)}]\delta(t-t_\ell)
 \end{equation}
 which transfer the momentum $\hbar \bm{k}_\ell^{(\alpha)}$ on branch $\alpha$ at time $t=t_\ell$ and imprint the laser phase $\varphi_\ell^{(\alpha)}$ evaluated at the time of the pulse on the wave packet.
In this article we assume that the unperturbed interferometer is closed, translating into the condition
\begin{equation}
\label{ClosedUnperturbed}
\hat{U}^{(l)\dagger}_0\hat{U}^{(u)}_0=\mathrm{e}^{\mathrm{i}\phi_0}\,,
\end{equation}
where $\hat{U}^{(\alpha)}_0$ is the time-evolution operator with respect to \Eq{UnperturbedHamiltonian}. The phase of the unperturbed interferometer $\phi_0$, is merely a $c$-number, implying perfect wave-packet overlap at the end of the  unperturbed interferometer sequence. However, the interferometer including the perturbation is in general not closed.

The time-evolution operator with respect to Hamiltonian \eqref{Hamiltonian} can be decomposed into
\begin{equation}
\label{InteractionPicture}
\hat{U}^{(\alpha)}=\hat{U}_0^{(\alpha)}\hat{U}_\mathrm{I}^{(\alpha)}
\end{equation}
by  transforming into the interaction picture with respect to $\hat{H}_0^{(\alpha)}$ so that the operator
\begin{equation}
\hat{U}_\mathrm{I}^{(\alpha)}=    \mathcal{T}\mathrm{exp}\left\{-\frac{\mathrm{i}}{\hbar}\int_{t_\mathrm{i}}^{t_\mathrm{d}} \!\mathrm{d}t\, \hat{V}_\mathrm{I}^{(\alpha)}(t)\right\}
\end{equation}
only includes the potential $\hat{V}_\mathrm{I}^{(\alpha)}(t)=V^{(\alpha)}(\hat{\bm{r}}^{(\alpha)}(t),t)$ which is  a function of $\hat{\bm{r}}^{(\alpha)}(t)=\hat{U}_0^{(\alpha)\dagger}\hat{\bm{r}}\hat{U}_0^{(\alpha)}$, the solution of the Heisenberg equations of motion generated by $\hat{H}_0^{(\alpha)}$.
An explicit expression for the solution
is straightforwardly  obtained  for our form of $\hat{H}_0^{(\alpha)}$,
resulting in
$\hat{\bm{r}}^{(\alpha)}(t)=\hat{\bm{r}}+\hat{\bm{p}}t/m+\tilde{\bm{r}}^{(\alpha)}_0(t)$. Here, $\tilde{\bm{r}}^{(\alpha)}_0(t)$ are the classical trajectories caused by the unperturbed Hamiltonian \eqref{UnperturbedHamiltonian}  with the initial conditions  $\tilde{\bm{r}}^{(\alpha)}_0(t_\mathrm{i})=0$ and  $\dot{\tilde{\bm{r}}}_0^{(\alpha)}(t_\mathrm{i})=0$ since the Schrödinger and Heisenberg picture coincide at $t=t_\mathrm{i}$.

\begin{figure}
	\begin{center}
		\includegraphics[width=\linewidth]{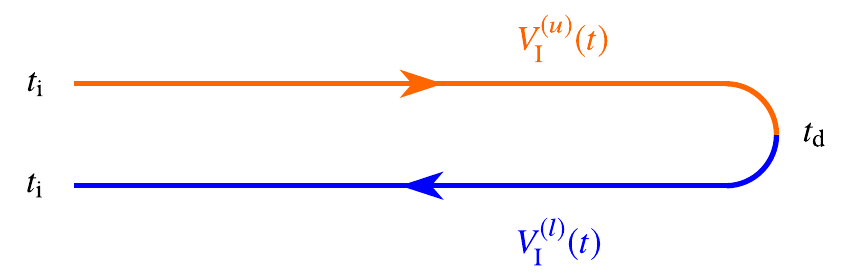}
		\caption{\textit{Time contour}. To merge the two time-evolution operators corresponding to the two branches of the interferometer, we introduce the time contour depicted in the figure over which all integrals extend. Starting at $t_\mathrm{i}$ the integrals run to the detection time $t_\mathrm{d}$ over the perturbation potential corresponding to the upper branch and subsequently return along the lower branch back to the initial time $t_\mathrm{i}$.}
		\label{fig:Contour}
	\end{center}
\end{figure}

Inserting the decomposition shown in \Eq{InteractionPicture}  into the overlap operator in  \Eq{OverlapOperator} then yields
\begin{equation}
\hat{U}^{(l)\dagger}\hat{U}^{(u)}=\hat{U}_\mathrm{I}^{(l)\dagger} \hat{U}^{(l)\dagger}_0\hat{U}^{(u)}_0\hat{U}_\mathrm{I}^{(u)}=\mathrm{e}^{\mathrm{i}\phi_0}\hat{U}_\mathrm{I}^{(l)\dagger} \hat{U}_\mathrm{I}^{(u)}\,.
\end{equation}
Recalling \Eq{ClosedUnperturbed}, we identified the phase of the closed unperturbed interferometer and moved the exponential $\mathrm{exp}(\mathrm{i}\phi_0)$ to the left as it is only a $c$-number. Writing the two interaction picture time-evolution operators explicitly
\begin{equation}
\hat{U}_\mathrm{I}^{(l)\dagger} \hat{U}_\mathrm{I}^{(u)}=\overline{\mathcal{T}}\mathrm{exp}\left\{\frac{\mathrm{i}}{\hbar}\int_{t_\mathrm{i}}^{t_\mathrm{d}} \!\mathrm{d}t\, \hat{V}_\mathrm{I}^{(l)}\right\}
 \mathcal{T}\mathrm{exp}\left\{-\frac{\mathrm{i}}{\hbar}\int_{t_\mathrm{i}}^{t_\mathrm{d}} \!\mathrm{d}t\, \hat{V}_\mathrm{I}^{(u)}\right\}\,,
 \end{equation}
 we note that the time-ordering operator $\mathcal{T}$  orders times (reading from right to the left) from the initial time $t_\mathrm{i}$ to the final time $t_\mathrm{d}$ while the anti-time-ordering operator $\overline{\mathcal{T}}$ orders from $t_\mathrm{d}$ back to $t_\mathrm{i}$. Thus, we merge the two time-evolution operators to one path-ordered exponential  by introducing the path-ordering operator $\mathcal{T}_p$ so that we obtain
 \begin{equation}
 \label{PathOrderedOverlap}
 \hat{U}^{(l)\dagger}\hat{U}^{(u)}=\mathrm{e}^{\mathrm{i}\phi_0}\mathcal{T}_p\,\mathrm{exp}\left\{-\frac{\mathrm{i}}{\hbar}\oint\! \mathrm{d} t\, \hat{V}_\mathrm{I}(t)\right\}\,.
\end{equation}
Resorting to the concept of path ordering, initially introduced by Schwinger and Keldysh \cite{PathOrdering1, PathOrdering2} in the context of thermal field theory, $\mathcal{T}_p$ orders time along the contour illustrated in \Fig{fig:Contour}. On this contour we  define $\hat{V}_\mathrm{I}(t)=\hat{V}_\mathrm{I}^{(u)}(t)$ for $t$ on the upper path and $\hat{V}_\mathrm{I}(t)=\hat{V}_\mathrm{I}^{(l)}(t)$ for $t$ on the lower path. 
With these definitions in mind, we disregard the explicit labeling $(\alpha )$ of the branch if not necessary.

In order  to associate the operator $\hat{\bm{r}}(t)$ with the unperturbed  trajectory of the atoms, 
the initial mean momentum and position   can be included  by introducing  $\bm{r}_0(t)=\tilde{\bm{r}}_0(t)+\langle \bm{\hat{r}}\rangle+\langle \bm{\hat{p}}\rangle t/m$ so that 
\begin{equation}
\label{Redefinition}
\hat{\bm{r}}(t)= \bm{r}_0(t)+\hat{\overline{\bm{r}}}(t)
\end{equation}
for each branch, where $\hat{\overline{\bm{r}}}(t)$ was defined in \Eq{WavePacket}.
Consequently, with \Eq{PathOrderedOverlap} the overlap operator can be written as a path-ordered exponential in which the perturbation potential is evaluated at the classical unperturbed trajectory plus an operator part with vanishing expectation value.
The expectation value
 $\langle \hat{\overline{{r}}}_j(t)^2 \rangle=\Delta r_j^2+\Delta p_j^2 t^2/m^2$, where $\Delta r_j$, $\Delta p_j$ are the initial position and momentum widths of the wave packet in the $j$th direction, is therefore a measure for the width of the expanding wave packet as long as  any distortion effects from the perturbation potential are negligible. Note that we disregarded for the moment possible initial correlations between $\hat{\bm{r}}$ and $\hat{\bm{p}}$.
 Thus, if the change of the potential over the size of the wave packet on each interferometer branch is sufficiently small, the Taylor expansion around the classical trajectory $\bm{r}_0(t)$
\begin{equation}
\label{TaylorExpansion}
\hat{V}_\mathrm{I}(t)=V+V_i\hat{\overline{{r}}}_i+\frac{1}{2}V_{ij}\hat{\overline{{r}}}_i\hat{\overline{{r}}}_j+...\,,
\end{equation}
 where
indices of the potential again denote derivatives,  accurately approximates the potential by taking into account only a few terms.
Note that we omitted the time dependence on the right-hand side for the sake of readability and again use summation convention. 

The overlap operator from  \Eq{PathOrderedOverlap} still contains the formal path-ordering operator $\mathcal{T}_p$, prohibiting any further manipulation of the contour-ordered exponential.
 To remove the former, we apply the Magnus expansion, detailed in \App{MagnusExpansion}, facilitating an exponential representation of a time-ordered exponential in terms of a formal series, namely
\begin{equation}
\label{Ansatz}
\hat{U}^{(l)\dagger}\hat{U}^{(u)}=\mathrm{e}^{\mathrm{i}\phi_0+\mathrm{i}\hat{\phi}}=\mathrm{exp}\left\{\mathrm{i}\sum_{n=0}^\infty \hat{\phi}_n\right\}\,,
\end{equation}
where $\hat{\phi}=\sum_{n=1}^\infty\hat{\phi}_n$. For a more convenient notation, the operator $\hat\phi_0$ is in fact defined as the phase of the unperturbed interferometer $\phi_0$, i.e. a $c$-number. The other contributions $\hat{\phi}_n $ for $n>0$ are determined by the Magnus expansion. In the appendix we provide the terms explicitly to third order; and in general $\hat{\phi}_n$ contains $n$ nested integrals along the time contour over nested commutators of order $n-1$ between the potential evaluated at different times.

The Magnus expansion therefore leads to an operator expansion in powers of the perturbation potential $V$. If small as defined in the following section, the
Magnus expansion can be truncated at  desired order and we have already succeeded in finding an approximate exponential
representation of the overlap operator. However, according to \Eq{OverlapOperator}, it is still necessary to evaluate the  expectation value of this operator.
In case of a path-independent harmonic potential it can be shown~\cite{Kleinert} that the exponent of the overlap operator depends only linearly on $\hat{\bm{r}}$ and $\hat{\bm{p}}$, making the calculation of the expectation value straightforward. However, in general such a simple representation does not exist and the overlap operator will contain various powers of $\hat{\bm{r}}$ and $\hat{\bm{p}}$. In this case, a suitable approach lies in the cumulant expansion, a formal series, casting the expectation value of an exponential operator into the form
\begin{equation}
\label{CumulantExpansionInText}
\langle\mathrm{e}^{\mathrm{i}\hat{\phi}}\rangle=\mathrm{exp}\left\{\sum_{n=1}^\infty\frac{\kappa_n}{n!}\right\}\,,
\end{equation}
where the cumulants $\kappa_n$, defined in \App{AppCumulants}, are functions of the first $n$ moments of $\hat{\phi}$.  
Consequently, the phase of an interferometer can  at least formally be expressed in powers of the perturbation potential.

Truncating the series at some desired order, however, requires a detailed assessment of  conditions characterizing the magnitude of the potential and the size of wave-packet effects. These conditions for the validity of our approach will be discussed in the subsequent section.

For  the moment let us assume that these conditions are satisfied, allowing a truncation of the Magnus and cumulant expansion at first order. Consequently, with the help of \Eq{Ansatz} and \Eq{cumphase} the phase is $\phi=\phi_0+\langle \hat{\phi}_1\rangle$. Inserting the Taylor polynomial of the potential up to quadratic order into  \Eq{FirstOrderMagnus} and recalling that $\langle\hat{\overline{\bm{r}}}\rangle=0$, we arrive at \Eq{simplifiedFormula}, the main result of our article.

\section{Conditions for validity}
\label{Validity conditions}
\begin{figure}
	\begin{center}
        \includegraphics[width=\linewidth]{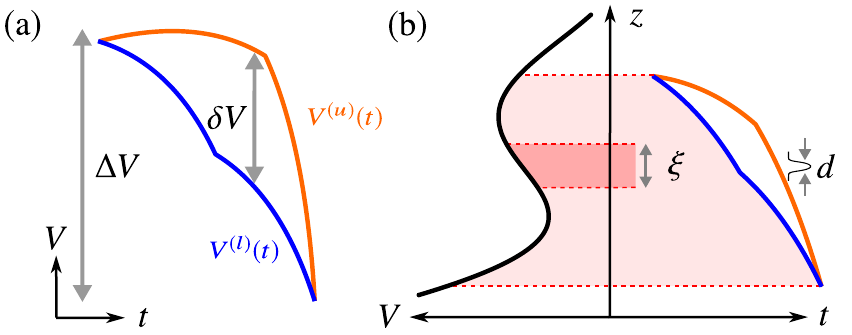}
		\caption{\textit{Comparison between different scales}. (a) Perturbation potential evaluated at the two unperturbed trajectories spanning the interferometer as a function of time (blue and orange curve). Whereas $\Delta V$ corresponds to the maximal potential difference probed by the atoms, $\delta V$ is the maximal potential difference between the branches of the interferometer, which is generally smaller.
		(b) The space-time diagram of a given interferometer sequence determines the domain of  the potential probed by the atoms (colored in bright red). The suppression factor of consecutive terms in the Magnus expansion as well as wave-packet effects are crucially dependent on the value of $\xi$ which is the length scale on which the potential changes. If we assign a wave number to the oscillatory behavior of the potential shown in the figure, $\xi$ is given by its inverse (colored in darker red). In contrast, in the case of a simple polynomial form of the potential, the value of $\xi$ would scale with the extent of the interferometer itself. A third length scale is determined by the characteristic size of the wave packet $d$.}
		\label{fig:Veranschaulichung}
	\end{center}
\end{figure}

We now derive conditions for the validity of the perturbative treatment presented in the section above.
First, we introduce a characteristic length scale $\xi$ with
\begin{equation}
\label{scaling}
    \Delta V\sim\xi^n \Delta V^{(n)}\,,
\end{equation}
where  $\Delta V$  is the difference between the extremal values of the potential  probed by the atoms over the course of the interferometer. Furthermore, 
$\Delta V^{(n)}$ is the typical value of the 
 $n$th derivative of the potential.
For example in case of a power-law dependence of the potential, $\xi$  corresponds to the size of the atomic fountain in which the experiment is performed while for oscillating potentials the value of $\xi$ can be much smaller. 
The parameters $\Delta V$ and $\xi$ are visualized in \Fig{fig:Veranschaulichung} together with further quantities defined below.
First, a perturbative treatment is only valid, if the deviation of the unperturbed trajectories caused by the perturbation potential is small compared  to the characteristic length scale $\xi$ on which the potential varies.
Therefore, identifying an acceleration $a\sim - \Delta  V/(m \xi)$ of the atom due to the perturbation potential, where  $\Delta V/\xi\sim \Delta V^{(1)}$, we require that the distance $a T^2$ is much smaller than the characteristic length $\xi$, that is
\begin{equation}
\label{epsilon}
    \epsilon=\frac{\Delta V T^2}{\xi^2m}\ll1\,,
\end{equation}
where $T$ denotes the characteristic interferometer time which can be chosen to be equal to  the interrogation time of the interferometer. As shown more rigorously in \App{Derivation of validity conditions}, the parameter $\epsilon$ constitutes the factor by which subsequent orders of the Magnus expansion are suppressed.

We now consider the leading-order correction to the unperturbed phase, that is the second term on the right-hand side of \Eq{simplifiedFormula}. This term can be estimated to be of the size 
\begin{equation}
\label{etadef}
    \eta=\frac{\delta V T}{\hbar}\,,
\end{equation}
where $\delta V$ is the maximal potential difference between the branches at one instance of time, which can be much smaller than $\Delta V$, the difference between the two extremal values of the potential probed by the atoms during the course of the interferometer.

In \Eq{TaylorExpansion} we Taylor expanded the potential about the unperturbed trajectory over the size of the wave packet represented by the operator $\hat{\overline{\bm{r}}}$.
Heuristically replacing the position operators by the wave-packet width $d$, the Taylor expansion can be truncated after a few terms if $\Delta V^{(n)}d^n\ll \Delta V$ which translates with the help of \Eq{scaling} into 
    \begin{equation}
\label{WavePacketCondition_a}
  d/\xi\ll1 .
\end{equation}
Finally, if this condition holds, the \emph{leading}-order operator-valued term in $\hat{\phi}$ is due to the \emph{first}-order term in the Taylor expansion of the potential, taking the form  $\oint\! \mathrm{d}t V_i\hat{\overline{r}}_i/\hbar$.
To guarantee the validity of the cumulant expansion, which is a function of the moments of $\hat{\phi}$, this term should be much smaller than unity.
This requirement can be expressed by
\begin{equation}
\label{WavePacketCondition_b}
  \eta d / \xi\ll1
\end{equation}
after again replacing the operator $\hat{\overline{\bm{r}}}$ by $d$, the integration in time by the characteristic interferometer time $T$, and by making use of \Eq{scaling}.
A more rigorous derivation of these conditions can be found in \App{Derivation of validity conditions}.

In table~\ref{table:Parameter scales} we give approximate values for $\xi$,  $\epsilon$, and $\eta d/\xi$ for  some experiments or recent proposals.
As shown in the  table, 
the accuracy achieved by truncation of the Magnus expansion at first order often is already sufficient  and our formalism is well suited to be applied to these situations.\\
\begin{table}
\caption{Explicit values  for $\xi$, $\epsilon$, $d/\xi$, and $\eta d/\xi$ for different sources of the perturbation. In the table we show that all these examples satisfy the conditions for the validity of our approach. The parameters $\Delta V$, $\delta V$ and $\xi$ used for the calculation of $\epsilon$ and $\eta$ are estimated from the potentials stated in the references. The width of the wave packet is assumed to be $d=50\,$\textmu m.\\}
\label{table:Parameter scales}
\centering
\begin{tabular}{ p{0.8cm}p{1.75cm}p{1.75cm}p{1.75cm}p{1.55cm} }
\hline\hline
  & gravity \newline gradients~\cite{Peters} & magnetic field\newline gradients~\cite{Magneticshielding} & blackbody \newline radiation~\cite{BBR1} & mass defect in quantum clocks~\cite{Redshift} \\
  \hline
    \vspace{-0.1cm}$\xi\;\,[m]$   & \vspace{-0.1cm}$0.1$    &\vspace{-0.1cm}$0.1$ & \vspace{-0.1cm} $0.01$ & \vspace{-0.1cm}  $10$\vspace{0.1cm}\\
  
   \vspace{-0.1cm}$\epsilon$   & \vspace{-0.1cm}$10^{-8}$    &\vspace{-0.1cm}$10^{-11}$ & \vspace{-0.1cm} $10^{-5}$ & \vspace{-0.1cm}  $10^{-12}$\vspace{0.1cm}\\
 \vspace{-0.1cm}$d/\xi$   &\vspace{-0.1cm} $5\cdot 10^{-4}$   &\vspace{-0.1cm} $5\cdot 10^{-4}$  &\vspace{-0.1cm}   $5\cdot 10^{-3}$&\vspace{-0.1cm}  $5\cdot 10^{-6}$\vspace{0.1cm}\\
 \vspace{-0.1cm}$\eta d/\xi$    &\vspace{-0.1cm} $10^{-4}$   &\vspace{-0.1cm} $10^{-9}$ &\vspace{-0.1cm}   $10^{-3}$ &\vspace{-0.1cm}   $10^{-9}$\vspace{0.1cm}\\
 \hline
 \hline
\end{tabular}
\end{table}
Obviously, a scaling such as in \Eq{scaling}   cannot be guaranteed in general. For this case we apply the Magnus expansion  to the Taylor polynomial of the perturbation potential explicitly in \App{Phase to second order} from which the scaling of phase shifts can be inferred.
However, for many applications a parameter $\xi$ does exist, satisfying \Eq{scaling} at least approximately, which is sufficient for an order-of-magnitude estimation of the size of  phase shifts beyond \Eq{simplifiedFormula}.

\section{Example: Gravitational potential}
\label{Gravitationalpotential}
In this section we illustrate the formalism derived in the previous sections by the example of an MZ-interferometry experiment conducted in the Newtonian gravitational potential of Earth
\begin{equation}
\label{PotentialEarth}
    V(\bm{r})=-G\frac{m M}{r}\,,
\end{equation}
where $G$ is Newton's constant and  $M$ is the mass of Earth.
Choosing the $z$ axis of a new coordinate system in direction of Earth's radius $R$, we  Taylor expand  the potential around a point on   Earth's surface in powers of $R^{-1}$.
This calculation yields 
\begin{equation}
\label{NewtonscheGravity}
    V(\bm{r})=mgz+\frac{1}{2}m\Gamma^{(1)}_{ij}{r}_i{r}_j+\frac{1}{6}m\Gamma^{(2)}_{ijl}{r}_i{r}_j{r}_l+\mathcal{O}(R^{-3})
\end{equation}
after omitting the irrelevant constant.
The only non-vanishing components of the fully symmetric first and second gravity-gradient tensors $\Gamma^{(1)}$ and $\Gamma^{(2)}$ then are given by
\begin{equation}
  \Gamma^{(1)}_{xx}=\Gamma^{(1)}_{yy}=\frac{g}{R},\quad\quad    \Gamma^{(1)}_{zz} =-2\frac{g}{R}\,,
\end{equation}
and (including all possible permutations)
\begin{equation}
 \Gamma^{(2)}_{xxz}=\Gamma^{(2)}_{yyz}=-3\frac{g}{R^2},\quad\quad \Gamma^{(2)}_{zzz}=6\frac{g}{R^2}\,,
\end{equation}
where we identified $g=G M/R^2$.
Phase corrections caused by  the first gradients (described by $\Gamma^{(1)}$)  have been calculated to all orders in $\Gamma^{(1)}$ \cite{Borde1, Kleinert}. However, note that phase contributions to second order in $\Gamma^{(1)}$ and to first order in $\Gamma^{(2)}$ both scale with $R^{-2}$ and may therefore be of the same size, making the calculation inconsistent when disregarding $\Gamma^{(2)}$.

In this example we  focus on  the second gravity gradients (described by $\Gamma^{(2)}$). Phase shifts from this contribution have been calculated before \cite{Hogan} but here we will additionally include the contribution of wave-packet effects. Furthermore, we stress how straightforward the calculation becomes with our formalism and put particular emphasis on the application of the conditions for the validity of a perturbative description.\\

\noindent
\textit{Check of conditions for validity}\\
 First we  check condition \eqref{epsilon} for the validity of the Magnus expansion.
Because of  the polynomial form of the potential, $\xi$ is given by the total extent of the interferometer. Assuming the recoil velocity  $v_\mathrm{r}=\hbar k/m$, where $\hbar k$ is the effective momentum transfer of the lasers as well as the initial velocity of the atoms $v_\mathrm{i}$ to be much smaller than $gT$, the size of the interferometer scales with $\xi=gT^2/2$ and we choose $\Delta V=m\Gamma^{(2)}\xi^3$, where $T$ is half of the interferometer time (see \Fig{fig:MZ}).
Consequently, with the help of \Eq{epsilon}
\begin{equation}
\epsilon=\Gamma^{(2)}gT^4\approx 10^{-12}
\end{equation}
for $T\approx1\,\mathrm{s}$. Consecutive terms in the Magnus expansion corresponding to the same power of the operators $\hat{\bm{r}}$ and $\hat{\bm{p}}$ are therefore suppressed by this  factor and a consideration to first order is sufficient.
Next we calculate the  potential difference between the interferometer branches by inserting the analytic expressions for the unperturbed trajectories from \Fig{fig:MZ}  into the potential so that
\begin{equation}
\delta V\sim m\Gamma^{(2)}[(v_\mathrm{r}T-gT^2/2)^3-(-gT^2/2)^3]\sim m\Gamma^{(2)}g^2v_\mathrm{r}T^5\,,
\end{equation}
where we replaced $t\sim T$. Consequently, the leading-order phase shift from the second gravity gradients is of the order of
\begin{equation}
    \eta=\Gamma^{(2)}\frac{mg^2v_\mathrm{r}T^6}{\hbar}=\Gamma^{(2)} k g^2 T^6\approx 10^{-4}\,,
\end{equation}
where, as the momentum transfer stems from a two-photon process, the value
$k=4\pi/(780\,\mathrm{nm})$ of the effective wave number  corresponds to twice the wave length of the  $\mathrm{D}_2$ line of rubidium $87$.

Finally, it is left to examine condition (\ref{WavePacketCondition_b}) for the validity of the cumulant expansion. Note that in principle one should  consider operator-valued contributions in the overlap operator from the first \textit{and} second gravity gradients  but  for the sake of a simple presentation we restrict the discussion to latter. Assuming a maximal size
$d=200\,$\textmu m of the wave packet, we find
\begin{equation}
\eta d/\xi\approx 10^{-9}
\end{equation}
so that \Eq{simplifiedFormula} can  be confidently applied.\\
\begin{figure}
	\begin{center}
        \includegraphics[width=\linewidth]{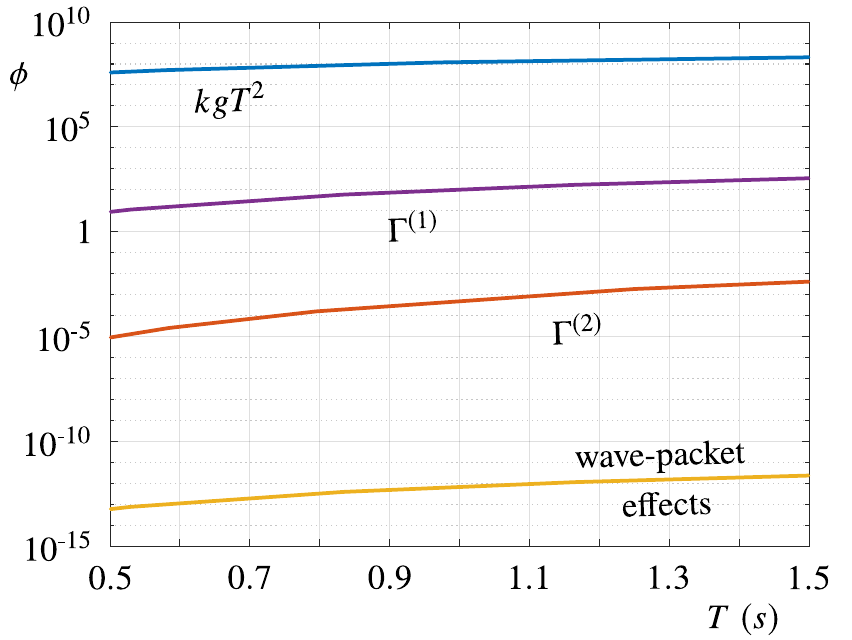}
		\caption{\textit{Plot of  phase shifts caused by the gravitational field of Earth.} As shown in the figure, the dominant contribution $kgT^2$ to the phase (blue line) is due to the unperturbed interferometer. The phase shift caused by the first gravity gradients \cite{Peters} scales with $R^{-1}$ (purple line).  Phase contributions from the second gravity gradients scale with $R^{-2}$ and can be divided into the first term inside the bracket in   \Eq{sizeofPhases} (orange line) and the second term which originates from wave-packet effects (yellow line).
		For this figure the parameters are: The radius of Earth is $R\simeq6\cdot 10^6\,\mathrm{m}$, the effective wave vector is given by $k=4\pi/(780\,\mathrm{nm})$, an interferometer time of $T=1\,\mathrm{s}$ is assumed,  the initial trapping frequency is $\omega=2\pi\cdot 60\,\mathrm{Hz}$, and a mass corresponding to rubidium $87$ is chosen. }
		\label{fig:Phasenplot}
	\end{center}
\end{figure}

\noindent
\textit{Calculation of phase}\\
We now consider a general  interferometer sequence, but for simplicity we  assume the laser pulses to be aligned with the direction of linear
gravity so that  the classical unperturbed trajectories take the form ${r}_{0j}(t)=z_0(t)\delta_{zj}$. Insertion of the perturbing potential into \Eq{simplifiedFormula} then yields for the phase shift  due to $\Gamma^{(2)}$
\begin{align}
\label{Phasecubic}
&\frac{1}{\hbar}\oint \mathrm{d} t\, V(t)=\frac{m}{6\hbar}\Gamma^{(2)}_{zzz}f_\phi\\
\intertext{and}
&\frac{1}{2\hbar}\oint\! \mathrm{d} t\, V_{ij}(t)\,\langle\hat{\overline{{r}}}_i(t)\hat{\overline{{r}}}_j(t)\rangle=\frac{1}{2\hbar}\Gamma^{(2)}_{zij}\Big[m\langle\hat{{r}}_i\hat{{r}}_j\rangle_cf_{rr}\\ 
\label{WavePacketCubic}
&\hspace{1.7cm}+\langle\hat{{r}}_i\hat{{p}}_j+\hat{{p}}_j\hat{{r}}_i\rangle_cf_{rp}+\frac{1}{m}\langle\hat{{p}}_i\hat{{p}}_j\rangle_cf_{pp}\Big]\,,
\end{align}
where we recalled \Eq{WavePacket} and defined the \textit{central expectation value} $\langle \cdot \rangle_c$ with respect to the initial state but displaced by $\langle\hat{\bm{r}}\rangle$ and $\langle\hat{\bm{p}}\rangle$ to the origin of phase space so that e.g.~$\langle\hat{{r}}_i\hat{{r}}_j\rangle_c=\langle(\hat{{r}}_i-\langle \hat{{r}}_i\rangle)(\hat{{r}}_j-\langle \hat{{r}}_j\rangle)\rangle$.
In \Eqs{Phasecubic} and \eqref{WavePacketCubic} we made furthermore use of the symmetry of $\Gamma^{(2)}$. The chosen interferometer sequence only enters the expression through the functions $f_\phi$, $f_{rr}$, $f_{rp}$ and $f_{pp}$  defined in \App{GravitationaPotential} and also explicitly evaluated  for the MZ geometry depicted in \Fig{fig:MZ}. In order to keep the expressions compact, we assume the initial position ${r}_{0j}(t_\mathrm{i})=z_\mathrm{i}\delta_{jz}$ without an initial velocity. If necessary, however, a more general calculation is straightforward.
As shown in the appendix, the phase then depends on the initial position  $z_\mathrm{i}$ of the state.
This result is a direct consequence of the operator-valued form of the overlap operator. In contrast, if the interferometer is closed, the overlap operator reduces to a $c$-number and its expectation value is independent of the wave function and therefore of the initial conditions.

To obtain compact explicit expressions for the phase shift due to wave-packet effects, we assume that the first laser pulse acts right after releasing the atoms from the trap.  Thus, assuming the initial state as the  ground state of a harmonic trap with frequency $\omega_i$ in the $i$th direction results in $\langle\hat{{r}}_i\hat{{r}}_j\rangle_c=\delta_{ij}\hbar/(2m\omega_i)$ and $\langle\hat{{p}}_i\hat{{p}}_j\rangle_c=\delta_{ij} \hbar m \omega_i/2$ as well as vanishing correlations between momentum and position operators. Specifying further $\omega\equiv\omega_x=\omega_y=2\omega_z$, the final result for the lowest-order correction due to the cubic potential is 
\begin{equation}
\label{sizeofPhases}
    \phi=\phi_0-\frac{g}{R^2}\left[\frac{m}{\hbar}f_\phi+\frac{v_\mathrm{r}T^2}{\omega}\left(\frac{3}{2}-\frac{7}{8}(\omega T)^2\right)\right]\,.
\end{equation}
Interestingly, if the atoms expanded out of a symmetric harmonic trap, the wave-packet induced contributions would  vanish.

In \Fig{fig:Phasenplot} we illustrate  \Eq{sizeofPhases} for reasonable experimental parameter values. In the figure we compare the phase of the unperturbed interferometer sequence (blue line) and the phase due to $\Gamma^{(1)}$  (purple line) \cite{Peters} to the phase shift calculated in \Eq{sizeofPhases} from the cubic contribution of the Taylor expansion (orange line).
The effect of phase shifts caused by $\Gamma^{(1)}$ are generally  relevant in state-of-the-art precision  measurements \cite{Peters}.
As a consequence, such phase contributions either have to be included into the analysis, or have to be compensated through differential schemes \cite{WEP3} (for example used for test of the weak equivalence principle) and mitigation techniques~\cite{Compensation1}.
Phase shifts originating from second gradients $\Gamma^{(2)}$, however, are of the order of magnitude to possibly limit future spaceborne missions if not appropriately accounted for.
Finally, the phase shift originating from different expansion dynamics along the branches (yellow line) is beyond any accessible value for light-pulse atom interferometric experiments in the mid future. 
Such phases, however, can be much larger when taking into account the inhomogeneous  gravitational field of the laboratory environment.

\section{Discussion and Conclusion}
Obviously, Hamiltonian \eqref{Hamiltonian} does not account for atom-atom interactions. However, state-of-the-art atom interferometers employ Bose-Einstein condensates as highly-coherent atom sources which are intrinsically interacting many-body systems. Nonetheless, numerical propagation of the initial mean-field state with the help of the Gross-Pitaevskii equation, including release from the trap and possibly magnetic lensing \cite{DeltaKick}, shows that due to the dynamical expansion, the strength of interactions quickly decreases. Thus, if the initial expansion time before the first laser pulse is sufficiently large, any further evolution will be accurately described by the Schrödinger equation and our formalism is valid from this instance of time. Then, the state right before the first laser pulse is used as input  \cite{OvercommingLoss} and the expectation value of the overlap operator is calculated with respect to this state. Even more if interactions are  negligible during the whole experiment, we include the time between release and first laser pulse into the unperturbed trajectories so that the influence of the perturbation on the wave packet during this initial expansion time is automatically accounted for. It then suffices to calculate expectations values with  respect to the ground state of the trap.

In our article we have restricted the discussion to perturbations which only depend on the position operator $\hat{\bm{r}}$. A generalization, however, of our results for the application to $\hat{\bm{p}}$-dependent perturbations, present for example in rotating frames, is straightforward. 

The application of the Magnus expansion to the overlap operator has resulted in nested contour integrals whoose  analytical evaluation becomes cumbersome for large orders of the expansion.
Even though, the integrals can be reordered \cite{Ufrecht}  to streamline analytical calculations, the loop structure of the integral is particularly useful in a numerical implementation.
Here, the perturbation potential (and its derivatives) are  discretized exactly on the time contour so that   a numerical integration algorithm  will automatically account for the loop properties of the integrals.

Our approach is applicable to a variety of situations with different sources of the perturbation, including e.g.~blackbody radiation, gravity gradients, inhomogeneities in the gravitational potential of the laboratory environment, magnetic field gradients, relativistic effects, violation parameters of the universality of free fall, and finite laser pulse lengths.

In this work we proposed a new perturbative tool to assess small phases due to spurious influences for light-pulse atom interferometers. Making use of a path-dependent description formalized by the introduction of the path-ordering operator,
we emphasized how the method solves the problems of previous results. Based on two formal series, the Magnus and cumulant expansion, we derived  \Eq{simplifiedFormula} for the leading-order phase shifts  originating from the perturbation including wave-packet effects and obtained detailed conditions for the validity of our approach which are stated in \Eq{epsilon}, \Eq{WavePacketCondition_a} and \Eq{WavePacketCondition_b}.
Finally, we commented on straightforward generalizations and the
numerical implementation in case  analytic calculations are not possible.

\section{Acknowledgement}
We thank F.~Di Pumpo, A.~Friedrich, A.~Roura, and W.~P.~Schleich for helpful discussions.
This work is supported by the German Aerospace Center (\href{http://dx.doi.org/10.13039/501100002946}{Deutsches Zentrum für Luft- und Raumfahrt}, DLR) with funds provided by the Federal Ministry for Economic Affairs and Energy (\href{http://dx.doi.org/10.13039/501100006360}{Bundesministerium f\"ur Wirtschaft und Energie}, BMWi) due to an enactment of the German Bundestag under Grant Nos. DLR~50WM1556 and 50WM1956.
We thank the Ministry of Science, Research and Art Baden-Württemberg (\href{http://dx.doi.org/10.13039/501100003542}{Ministerium f\"ur Wissenschaft, Forschung und Kunst Baden-Württemberg}) for financially supporting the work of IQ$^\mathrm{ST}$.

\appendix
\section{Magnus expansion}
\label{MagnusExpansion}
The Magnus expansion \cite{ Magnus2, Magnus1,Ufrecht} is a formal series for the exponential representation of a time-ordered exponential
\begin{equation}
\label{NormalMagnusExpansion}
\hat{U}=\mathcal{T} \mathrm{exp}\left\{-\frac{\mathrm{i}}{\hbar}\int_{t_\mathrm{i}}^{t_\mathrm{d}} \!\mathrm{d}t\, \hat{H}(t)   \right\}
=\mathrm{exp}\left\{\mathrm{i}\sum_{n=1}^\infty \hat{\phi}_n   \right\}\,,
\end{equation}
where $\hat{H}(t)$ is a time-dependent Hamiltonian.
Applied to the path-ordered exponential in \Eq{PathOrderedOverlap},  we obtain for the first three elements of the series in \Eq{Ansatz}
\begin{align}
\label{FirstOrderMagnus}
\hat{\phi}_1&=-\frac{1}{\hbar}\oint \!\mathrm{d}t\, \hat{V}_\mathrm{I}(t)\\
\label{SecondOrderMagnus}
\hat{\phi}_2&=\frac{\mathrm{i}}{2\hbar^2}\oint\!\mathrm{d}t\!\oint^{t} \!\mathrm{d}t^\prime\, [\hat{V}_\mathrm{I}(t),\hat{V}_\mathrm{I}(t^\prime)]\\
\hat{\phi}_3&=\frac{1}{6\hbar^3}\oint \!\mathrm{d}t\!\oint^{t} \!\mathrm{d}t^\prime\!\oint^{t^\prime} \!\mathrm{d}t^{\prime\prime}\, \Big( [\hat{V}_\mathrm{I}(t),[\hat{V}_\mathrm{I}(t^\prime),\hat{V}_\mathrm{I}(t^{\prime\prime})]]\\
&\quad\quad\quad\quad\quad\quad\quad\quad\quad\quad\quad\quad+[\hat{V}_\mathrm{I}(t^{\prime\prime}),[\hat{V}_\mathrm{I}(t^\prime),\hat{V}_\mathrm{I}(t)]] \Big)\,.
\end{align}
In general,  $\hat{\phi}_n$ consists of $n$ nested integrals over  $(n-1)$th-order commutators between the potential evaluated at different times. The  perturbation potential is a function of the solution of the Heisenberg equations of motion $\hat{\bm{r}}(t)$ generated by the unperturbed Hamiltonian.

\section{ Cumulant expansion}
\label{AppCumulants}
The cumulant expansion \cite{Cumulants2, Cumulant} is a formal series for an exponential representation of the expectation value of  an exponential operator which is defined by 
\begin{equation}
\label{CumulantExpansion}
\langle\mathrm{e}^{\mathrm{i}\hat{\phi}\zeta}\rangle=\mathrm{exp}\left\{\sum_{n=1}^\infty\frac{\kappa_n}{n!}\zeta^n\right\}\,,
\end{equation}
where one introduces a formal expansion parameter $\zeta$, which is set to unity after the calculation, and the coefficients $\kappa_n$ are referred to as \textit{cumulants}. By taking the logarithm on both sides, we find the definition of the cumulants as
\begin{equation}
\kappa_n=\frac{\mathrm{d}^n}{\mathrm{d}\zeta^n}\mathrm{ln}\langle \mathrm{e}^{\mathrm{i} \hat{\phi} \zeta }  \rangle \Big|_{\zeta=0}\,,
\end{equation}
where the $n$th cumulant is function of the first $n$ moments of $\hat{\phi}$.
Here, we state explicitly the first three cumulants
\begin{align}
\kappa_1&=\mathrm{i}\langle \hat{\phi}\rangle\\
\kappa_2&=-\langle \hat{\phi}^2\rangle+\langle \hat{\phi}\rangle^2\\
\kappa_3&=-\mathrm{i}[\langle \hat{\phi}^3\rangle-3\langle \hat{\phi}^2\rangle\langle \hat{\phi}\rangle+2\langle \hat{\phi}\rangle^3]\,.
\end{align}
Since $\hat{\phi}=\sum_{n=1}^\infty \hat{\phi}_n$  is calculated from the overlap operator by Magnus expansion and is therefore  Hermitian, we separate \Eq{CumulantExpansion} into phase and amplitude. By comparing to \Eq{OverlapOperator}, we find the phase $\phi$ of the interferometer 
\begin{equation}
\label{cumphase}
 \phi=  \phi_0 +\langle \hat{\phi}\rangle-\frac{1}{6}[\langle \hat{\phi}^3\rangle-3\langle \hat{\phi}\rangle^2\langle \hat{\phi}\rangle+2\langle \hat{\phi}\rangle^3]+...\,,
 \end{equation}
where we included the phase of the unperturbed interferometer $\phi_0$ from \Eq{Ansatz}, and the contrast $C$ is
 \begin{equation}
 \label{Contrastdrop}
 \mathrm{ln}C=-\frac{1}{2}(\langle \hat{\phi}^2\rangle-\langle \hat{\phi}\rangle^2)+...\,.
\end{equation}

\section{Phase to second order}
\label{Phase to second order}
In this appendix we apply the Magnus expansion to second order to the overlap operator.
To this end, we insert the Taylor series of the potential into \Eq{Ansatz},
where $\hat{\phi}_n$ is determined by the Magnus expansion. Making use of the commutator
\begin{equation}
\label{commutator}
    [\hat{\overline{{r}}}_i(t)\,,\,\hat{\overline{{r}}}_j(t^\prime)]=\frac{\mathrm{i}\hbar}{m}(t^\prime-t)\delta_{ij}\,,
\end{equation}
where $\delta_{ij}$ is the Kronecker symbol,
we obtain  with the help of \Eqs{FirstOrderMagnus} and \eqref{SecondOrderMagnus}
	\begin{align}
	\hat{\phi}_0=&\phi_0\\
	\hat{\phi}_1=&-\frac{1}{\hbar}\oint\! \mathrm{d} t\, \bigg\{V+V_i\hat{\overline{{r}}}_i  +\frac{1}{2}V_{ij}\hat{\overline{{r}}}_i\hat{\overline{{r}}}_j+\frac{1}{6}V_{ijk}\hat{\overline{{r}}}_i\hat{\overline{{r}}}_j\hat{\overline{{r}}}_k\\
	&\quad\quad\quad\quad\quad+\frac{1}{24}V_{ijkl}\hat{\overline{{r}}}_i\hat{\overline{{r}}}_j\hat{\overline{{r}}}_k\hat{\overline{{r}}}_l+... \bigg\}\\
	\hat{\phi}_2=&-\frac{1}{2\hbar m}\oint\! \mathrm{d} t\,                                \oint^t\!\!\! \mathrm{d} t^\prime\,(t^\prime-t)\bigg\{ V_iV_i^\prime+V_iV_{ij}^\prime\,\hat{\overline{{r}}}_j^\prime +V_i^\prime V_{ij}\,\hat{\overline{{r}}}_j\\
	&\quad\quad\quad\quad\quad+\frac{1}{2}V_i^\prime V_{ijk}\,\hat{\overline{{r}}}_j\,\hat{\overline{{r}}}_k+\frac{1}{2}V_{ik}V_{kj}^\prime\left(\hat{\overline{{r}}}_j^\prime\,\hat{\overline{{r}}}_i+\hat{\overline{{r}}}_i\,\hat{\overline{{r}}}_j^\prime\right)\\
	&\quad\quad\quad\quad\quad+\frac{1}{2}V_iV_{ijk}^\prime\,\hat{\overline{{r}}}_j^\prime\,\hat{\overline{{r}}}_k^\prime +...                 \bigg\}\,.
	\end{align}
	Note that all quantities depend on time $t$ or $t^\prime$. If dependent on the latter, this dependence is abbreviated by a prime on the respective quantity.

\section{Derivation of conditions for validity}
\label{Derivation of validity conditions}
In order to complement the validity discussion of \Secref{Validity conditions}, we proceed in two steps. First, we derive the factor $\epsilon$  by which subsequent terms  in the Magnus expansion are suppressed. Second, we investigate the scaling of different orders in the cumulant expansion.

For the sake of simplicity, in the following  we  choose one typical direction  $x$ and suppress the time dependence of $\hat{\overline{x}}(t)$ except when appearing in the commutator.
Within this simplification we replace the potential and its derivatives in \Eq{TaylorExpansion} by their typical size and use \Eq{scaling} to find $\hat{V}_\mathrm{I}\sim\Delta V\sum_k \hat{\overline{x}}^k/\xi^k$.
With the help of this form of the potential,  the commutator  becomes
\begin{equation}
    [\hat{V}_\mathrm{I}(t),\hat{V}_\mathrm{I}(t^\prime)]\sim \frac{\Delta V^2\hbar T}{m\xi^2}\sum_k  \frac{\hat{\overline{x}}^k}{\xi^k}\,,
\end{equation}
where we replaced the time difference in  \Eq{commutator} by the characteristic interferometer time $T$ and suppressed any numerical factors. This result is easily generalized to the $(n-1)$th order nested commutator (which contains the potential evaluated at $n$ different times) as
\begin{equation}
\label{Zwischenschritt}
    [\hat{V}_\mathrm{I}(t),[\hat{V}_\mathrm{I}(t^\prime),[...]]]_{n-1}\sim \frac{\Delta V\hbar^{n-1} }{T^{n-1}}\epsilon^{n-1}\sum_k  \frac{\hat{\overline{x}}^k}{\xi^k}\,,
\end{equation}
where $\epsilon$ is given in \Eq{epsilon}. According to \App{MagnusExpansion} the $n$th-order term  $\hat{\phi}_n$ of the Magnus expansion contains $(n-1)$th-order commutators, a factor $1/\hbar^n$, and $n$  integrals over time which we replace by $T^n$. Hence, together with \Eq{Zwischenschritt} one again obtains an  infinite series
\begin{equation}
\label{scalingphi}
    \hat{\phi}_n\sim\sum_k c_{kn}\hat{\overline{x}}^k\,,
\end{equation}
where 
\begin{equation}
\label{scalingc}
  c_{kn}  \sim \frac{\eta\epsilon^{n-1}}{\xi^k}
\end{equation}
and $\eta$ was defined in \Eq{etadef}.
Note that, since the final integral in the nested sequence extends along the whole contour, we replaced one factor of $\Delta V$ by the maximal potential difference over the separation of the branches $\delta V$. 
Consequently, consecutive terms in the Magnus expansion corresponding to the same power of $\hat{\overline{x}}$ are suppressed by $\epsilon$ independently of the power.
Hence, if $\epsilon\ll1$,  the Magnus expansion can be truncated as the prefactors $c_{kn}$ quickly decrease order by order of $n$. 

After performing the Magnus expansion, it remains to calculate the  expectation value of the overlap operator. Because this is in general not possible in an exact manner, we resort to the cumulant expansion for which expectation values of powers of $\hat{\overline{{r}}}_j$ at different times have to be evaluated.
To estimate the size of such expectation values  independently of the explicit form of the initial wave function, we truncate the corresponding probability density outside some region with characteristic width $d$, where the probability to find a particle is vanishing. This approximation will allow us to express any moment in terms of the finite width $d$ of the wave function. Note that calculating phase and contrast with the help of the Magnus and cumulant expansions  might lead to divergent series but there will exist a finite number of terms after which truncating the formal series leads to the best approximation in the spirit of an asymptotic expansion.

In order to estimate the expectation value of powers of $\hat{\overline{{r}}}_j(t)$, we define the centered wave function $\ket{\psi_c}$, the initial state displaced by $\langle \hat{\bm{r}} \rangle$ and $\langle \hat{\bm{p}} \rangle$ to the origin of phase space. Equally, evolving with  the free time-evolution operator $\hat{\mathcal{U}}(t)=\mathrm{exp}[-\mathrm{i}{\hat{\bm{p}}^2}t/(2m\hbar)]$,  the freely expanding centered wave function is denoted by
$\ket{\psi_c(t)}$. Thus,
\begin{align}
\langle \hat{\overline{x}}^k(t) \rangle&=\bra{\psi_c} (\hat{x}+\hat{p}t/m)^k \ket{\psi_c}\\
&=
\bra{\psi_c} \hat{\mathcal{U}}^\dagger(t) \hat{x}^k \hat{\mathcal{U}}(t) \ket{\psi_c}=\bra{\psi_c(t)}  \hat{x}^k  \ket{\psi_c(t)}\\
&=\int\! \mathrm{d}^3\bm{r} |\psi_c(\bm{r},t)|^2 x^k=x_d(t)^k\sim d^k\,.
\end{align}
First, we removed the expectation values $\langle\hat{\bm{r}}\rangle$  and $\langle\hat{\bm{p}}\rangle$ appearing in the definition of 
$\hat{\overline{\bm{r}}}(t)$, see \Eq{WavePacket},
by calculating the expectation value with respect to  centered wave function rather than the actual initial state.
In the third line we used the mean-value theorem of integration \cite{MeanValue}  to find the number $x_d(t)$ within the set where the wave function is nonvanishing, subsequently estimated by the maximal size $d$ of the wave function. 
Thus, the expectation value of any power of $\hat{\overline{x}}(t)$ can indeed be expressed in terms of the width $d$.

However, expectation values of the commutators in the Magnus expansion or higher-order terms of the cumulant expansion involve products between powers of $\hat{\overline{x}}$ evaluated at different times.
The expectation value of such expressions can be calculated for example in Wigner phase space where one has to take care of the correct operator ordering  \cite{WignerSchleich, WignerCase}. Nevertheless, the scaling with the size of the wave packet remains similar so that we will e.g.~assume
\begin{equation}
\langle \hat{\overline{x}}^l(t) \hat{\overline{x}}^k(t^\prime) \rangle\sim d^{l+k}\,.
\end{equation}
With the help of \Eqs{scalingphi} and \eqref{scalingc} we therefore find 
\begin{equation}
\label{ExpectationScaling}
    \langle\hat{\phi}_n\rangle \sim\eta\epsilon^{n-1}\sum_k \left(\frac{d}{\xi}\right)^k
\end{equation}
for the expectation value of the $n$th order of the Magnus expansion.

We now investigate the behavior of the cumulant expansion by considering only the dominant operator-valued term in the overlap operator which is due to 
$n=1$ and $k=1$ in \Eq{scalingphi} provided $\epsilon\ll1$  and $d/\xi\ll1$.
As explained in \App{AppCumulants} the $l$th order of the cumulant expansion is a function of the first $l$ moments which consequently scales as $(\eta d/\xi)^l$ and we therefore require
\begin{equation}
\label{UnimpCond}
    \eta\frac{d}{\xi}\ll1\,,
\end{equation}
which is condition (\ref{WavePacketCondition_b}).
If satisfied, we also truncate the cumulant expansion at first order and obtain (considering only terms up to harmonic order in the Taylor expansion of the potential) 
\Eq{simplifiedFormula}, the main result of this article after recalling $\langle \hat{\overline{\bm{r}}}(t)\rangle=0$.\\

\section{Gravitational potential}
In this appendix we give the explicit form of the functions $f_\phi$, $f_{rr}$, $f_{rp}$ and $f_{pp}$ defined in \Secref{Gravitationalpotential} to calculate the phase and wave-packet effects arising from  $\Gamma^{(2)}$.
For an arbitrary pulse sequence encoded in the unperturbed trajectory $z_0(t)$, the coefficients take the form
\label{GravitationaPotential}
\begin{align}
f_\phi=&\oint\! \mathrm{d} t\, z_0(t)^3, &
f_{rr}=&\oint\! \mathrm{d} t\, z_0(t),\\
f_{rp}=&\oint\! \mathrm{d} t\, z_0(t)t,  \hspace{.65cm} \text{and}
&f_{pp}=&\oint\! \mathrm{d} t\, z_0(t)t^2.
\end{align}
Using the explicit form of $z_0(t)$ for the MZ interferometer sequence shown in \Fig{fig:MZ} with the initial conditions ${r}_{0j}(t_\mathrm{i})=z_\mathrm{i}\delta_{jz}$ and $\dot{r}_{0j}(t_\mathrm{i})=0$, we therefore find 
\begin{align}
f_\phi&=31 g^2 v_\mathrm{r}T^6/20-v_\mathrm{r} g T^4(14z_\mathrm{i}+9 v_\mathrm{r}T)/4 \\&\phantom{=}+v_\mathrm{r} T^2 (v_\mathrm{r}^2T^2+3v_\mathrm{r}T z_\mathrm{i}+3 z_\mathrm{i}^2)\\
f_{rr}&=v_\mathrm{r}T^2,\hspace{.2cm}
f_{rp}=v_\mathrm{r}T^3,\hspace{.2cm} \text{and}\hspace{.2cm}
f_{pp}=\frac{7}{6}v_\mathrm{r}T^4\,.
\end{align}

<AY

\end{document}